\documentclass{article}
\usepackage{arxiv}
\usepackage[utf8]{inputenc} 
\usepackage[T1]{fontenc}    
\usepackage{hyperref}       
\usepackage{url}            
\usepackage{booktabs}       
\usepackage{amsfonts}       
\usepackage{nicefrac}       
\usepackage{microtype}      
\usepackage{lipsum}
\usepackage{txfonts}
\usepackage{xcolor}
\usepackage{graphicx}
\usepackage{dcolumn}
\usepackage{bm}
\title{Quantum Travel Time and Tunnel Ionization Times of Atoms}
\author{
  Durmu{\c s} Demir\\
  Faculty of Engineering and Natural Sciences\\
  Sabanc{\i} University\\
  34956 Tuzla, {\.I}stanbul, Turkey\\
  \texttt{durmus.demir@sabanciuniv.edu}\\
\And
  Serkan Pa{\c c}al \\
  Department of Physics\\
  {\.I}zmir Institute of Technology\\
  35430 Urla, {\.I}zmir, Turkey  \\
  \texttt{serkanpacal@iyte.edu.tr}
  }
\begin{document}
\maketitle
\date{\today}

\begin{abstract}
Time it takes to travel from one position to another, devoid of any quantum mechanical description,  has been modeled variously, especially for quantum tunneling. The model time, if universally valid, must be subluminal, must hold everywhere (inside and outside the tunneling region), must comprise interference effects, and must have a sensible classical limit. Here we show that the quantum travel time, hypothesized to emerge with the state vector, is a function of the probability density and probability current such that all the criteria above are fulfilled. We compute it inside and outside a rectangular potential barrier and find physically sensible results. Moreover, we contrast it with recent ionization time measurements of the $\rm He$ as well as the $\rm Ar$ and $\rm Kr$ atoms, and find good agreement with data. The quantum travel time holds good for stationary systems, and can have applications in numerous tunneling-driven phenomena. 
\end{abstract}

\section{\label{sec:level1} Introduction}
Tunneling, transport of subatomic particles through the regions of
space forbidden to classical motion, is a pure quantum phenomenon.
It is a ubiquitous effect that underlies
numerous physical \cite{phys}, chemical \cite{chem}, biological
\cite{bio} and technological phenomena \cite{roy}.

Tunneling time, the time elapsed during the tunneling process, is
crucial for determining reaction speeds of tunneling-enabled rare
processes, which range from nuclear fusion \cite{gamow} to quantum 
annealing \cite{annealing-comp}. In fact, with the advent of strong laser ionization experiments \cite{exp0, exp1,exp11,exp12,exp13}, it has now become  possible to measure time of tunneling \cite{exp2,exp2-new}, where certain metrological problems \cite{exp-problem1,Torlina} with the detection of the tunneling particle were shown
to be surmountable \cite{exp-resolve1, exp-resolve2}. Strong laser
fields enable electrons to tunnel out of atoms, where the potential
barrier formed forms a testbed for models of tunneling time
\cite{keldysh,keldysh2}. In fact, recent single-electron ionization time measurements on $\rm He$ \cite{Maurer} (see also the more recent analysis \cite{Maurer2}), and $\rm Ar$ and $\rm Kr$ \cite{Yakaboylu} have shown that tunneling takes a finite time (via study of the $\rm Ar$ and $\rm Kr$ ionizations in \cite{Yakaboylu}). These experiments are sensitive to tunneling times ${\mathcal{O}}(100\ {\rm as})$ and this precision is sufficient to test various  tunneling time models. It is difficult to contrast theory and experiment, however. First, construction of 
the ionization potential in multi-electron atoms is highly complicated though, in this context, single-active-electron (SAE) approximation \cite{GSZ,Tong-lin, Muller} gives a satisfactory framework. (In our analyses below, we adopt the SAE potential given in \cite{Muller, Zhang}.) Second, stationarity of tunneling process brings limitations like, for example, the ionizing laser field must be sufficiently static (period of the laser field must be much larger than the tunneling time). Stationarity poses also a conceptual problem in that one must be able to set up a time measure (tunneling time) for a stationary process. Stationarity poses yet another problem in that potential loses its static nature in a duration about the laser period after the completion of tunneling process. (Sec. II and IV below take into account all these critical points.)

The time, not only the tunneling time, is an intricate concept in quantum theory.  The problem is that, in quantum theory, time is not a dynamical variable representable by an operator. It is not a  measurable quantity. It therefore is model-dependent and depends on the kinetic theory set forth for the tunneling process. The literature consists of various time definitions, as reviewed in
\cite{time-review,time-review2,time-review3,time-review4}. They
include traversal time through modulated barriers
\cite{buttiker-landauer, buttiker-landauer2, buttiker-landauer3,
japan-rec}, spin precession time \cite{larmour, larmour4,larmour5},
flux-flux correlation duration \cite{pollak-miller}, phase
time \cite{wigner,wigner2,wigner3}, and Feynman path
integral (FPI) averaging of the classical time
\cite{times-are-averages,times-are-averages2,times-are-averages3}.
Some of them are complex, some are difficult to associate with
tunneling process, and some suffer from superluminality \cite{superluminal,emt}. More importantly, they (excepting FPI time with experiment-driven coarse-grained paths) fail to explain the
experimental data, as was comparatively analysed and experimented in \cite{Maurer}. Nevertheless,  two recent time definitions, the entropic tunneling time \cite{demir-guner} and uncertainty-based tunneling time \cite{kullie}, are subliminal and agree well with $\rm He$ ionization data.  

In the present work, we propose and study a new time model which holds both {\it inside} and {\it outside} the potential barrier. This we do by structuring time as a function of the position of the particle (through its wavefunction), and guiding it with the Schr{\"o}dinger equation. Our definition, which we call quantum travel time (QTT), differs from those in the literature by its suitability for stationary processes like tunneling (Sec. II), its comprehensiveness for reflected and transmitted particles (Sec. III), its capability to hold everywhere (Sec. II and III), and its compatibility with experiment (Sec. IV). Indeed, in Sec. II below state the QTT. In Sec. III we study tunneling through a rectangular potential barrier as an illustrative example and as a testbed for physical consistency of QTT.  In Sec. IV, we contrast QTT with the experimental data on $\rm He$ \cite{Maurer} and $\rm Ar$ and $\rm Kr$ \cite{Yakaboylu} ionization times, and find fairly good agreement. In Sec. V we conclude. 

\section{Quantum Travel Time}
Our approach to time is novel in that it covers time elapsed both inside and outside the tunneling region. It is inspired by time in quantum gravity. More precisely, it uses  timelessness of the Hamiltonian general relativity (GR) \cite{adm, dewitt} as the starting point, with no real involvement of classical or quantized GR in its construction. The GR is timeless  because it is background independent  \cite{bkg}. Indeed, the Hamiltonian of 3-metrics $H$ vanishes identically
\begin{eqnarray}
\label{ham}
H = 0 
\end{eqnarray}
as a dynamical constraint. It means that dynamical variables in phase space are all time-independent.  Everything is frozen. 

The constraint (\ref{ham}), upon quantization, leads to  the Wheeler-DeWitt equation 
\begin{eqnarray}
\label{wdw}
{\hat{H}} |\psi\rangle = 0 
\end{eqnarray}
as a condition \cite{dewitt} on the state vector $|\psi\rangle$. This equation can be taken to imply, as in the classical theory, a strictly time-independent state vector. This, however, is not the only way. Indeed, the same equation  can interpreted as the Schr{\"o}dinger equation for the stationary state 
\begin{eqnarray}
\label{state-vector-time}
|\Psi(t)\rangle = e^{- i E t} |\psi\rangle
\end{eqnarray}
with vanishing energy ($E=0$). This reinterpretation makes a case that times does actually exist but is erased by vanishing energy. Unlike the classical Hamiltonian GR in which no dynamical variable possesses explicit time dependence, the quantized GR allows the possibility that timelessness is a result of the energylessness.  In fact, if the quantum gravitational system gets excited by some interactions (with matter, for instance) then time should emerge as in (\ref{state-vector-time}) thanks to non-vanishing  $E$.  In essence, the time $t$ should be a quantum property as it emerges together with the state vector $|\psi\rangle$. It is, however, not possible to disentangle it through (\ref{wdw}) due to vanishing of energy in quantum gravity \cite{dewitt,time-review}. It needs be structured separately.

 If time is to gain an observable status it must be related to observable quantities (position, momentum, energy, $\cdots$) in a measurable way. In this regard, the canonical quantum gravity setup in (\ref{wdw}), with the state vector (\ref{state-vector-time}), gives ground for a possible realization of observable time. It gives because $t$ and $|\psi\rangle$ are born together and $t$, as a c-number, must be related to the wavefunction $\psi(\vec{x},t)=\langle \vec{x}|\Psi(t)\rangle$ as a functional relation $t=t(\psi({\vec{x}}))$. Then, the trivial relation $dt/dt=1$ leads to $(d\vec{x}/dt)\cdot {\vec{\nabla}}t = 1$, which we generalize to quantum dynamics as  
\begin{eqnarray}
\label{time-eq}
{{\vec{J}}_{\rightsquigarrow}} \cdot \vec{\nabla}t = \rho -\rho_{\leftsquigarrow}
\end{eqnarray}
in terms of the probability current 
\begin{eqnarray}
{{\vec{J}}_{\rightsquigarrow}}=\frac{\hbar}{m}\Im\left[\psi^\dagger_{\rightsquigarrow} {\vec{\nabla}} \psi_{\rightsquigarrow}\right]
\end{eqnarray}
and the probability densities
\begin{eqnarray}
\rho_{\rightsquigarrow}=\left(\psi_{\rightsquigarrow} + \psi_{\leftsquigarrow}\right)^\dagger \left(\psi_{\rightsquigarrow} + \psi_{\leftsquigarrow}\right)\,,\; \rho_{\leftrightsquigarrow}=\psi^\dagger_{\leftsquigarrow} \psi_{\leftsquigarrow}
\end{eqnarray}
such that $\psi_{\rightsquigarrow}$ is the wavefunction propagating in the $\rightsquigarrow$ direction, that is, ${\vec{\nabla}} \psi_{\rightsquigarrow} \propto {\vec{p}}_{\rightsquigarrow} \psi_{\rightsquigarrow}$ (the momentum ${\vec{p}}_{\rightsquigarrow}$ along $\rightsquigarrow$  can be local or global, as was detailed in \cite{demir-guner}). The equation (\ref{time-eq}), resulting from quantum generalization of $d\vec{x}/dt$, is recognized to resemble the guiding equation in Bohmian mechanics \cite{dbb}. This reversible relationship lays the foundation for a proper formulation of time in quantum theory. Indeed, equation (\ref{time-eq}) governs how the time $t=t(\psi({\vec{x}}))$ emerges along with $\psi(\vec{x})$ not how the trajectory $\vec{x}(t)$ actualizes to kill the probabilistic nature of quantum behavior \cite{dbb}. The time $t=t(\psi({\vec{x}}))$, direct line integral over ${{\vec{J}}^{-1}_{\rightsquigarrow}} \left(\rho_{\rightsquigarrow}+\rho_{\leftrightsquigarrow}\right)$, makes sense only as temporal separations between the points $\vec{x}_a, \vec{x}_b, \cdots$ at which position measurements $a, b, \cdots$ are made. Hereon, we call the time definition (\ref{time-eq}) as quantum travel time (QTT) to emphasize the fact that particle ``travels" rather than traverses in face of forward and backward probability currents. 

In a one-dimensional setting (a characteristic feature of tunneling), time it takes to get from position $a$ to position $b$ takes the form
\begin{eqnarray}
\label{time-2}
(\Delta t)_{ba} = \int_{a}^{b} \frac{\left(\rho -\rho_{a\leftsquigarrow b}\right)}{J_{a\rightsquigarrow b}} d\ell_{ab}
\end{eqnarray}
in which the line element $d\ell_{ab}$ is directed from $a$ to $b$ (both lying along $\rightsquigarrow$ direction). Our time definition should not be confused with dwell time \cite{larmour}, which focuses on total probability $\rho$ and total current $|\vec{J}|$ to determine duration of scattering irrespective of if the particle is reflected or transmitted \cite{Dwellwkb}. Our definition in (\ref{time-2}), the QTT, is not a dwell time but a traversal time.  (The dwell time turns out to be superluminal as discussed in the next section, with the example of rectangular potential barrier.)

In the following we will use the time formula (\ref{time-2}) to compute time elapsed during potential scatterings. We first apply it to scattering form a rectangular potential barrier (in Sec. III below). Next, we apply it to single electron ionization from $\rm He$, $\rm Ar$ and $\rm Kr$ atoms, and contrast the results with the available experimental data (in Sec. IV). Our time definition naturally applies to stationary systems and, as will be seen in the sequel, it yields physically sensible results and shows satisfactory agreement with data.

\section{QTT For Rectangular Potential Barrier}
In this section we apply the time formula (\ref{time-2}) to rectangular potential scattering by considering the incident, reflected and transmitted waves separately. 

\begin{figure}[ht]
\centering
\includegraphics[width=8.6cm]{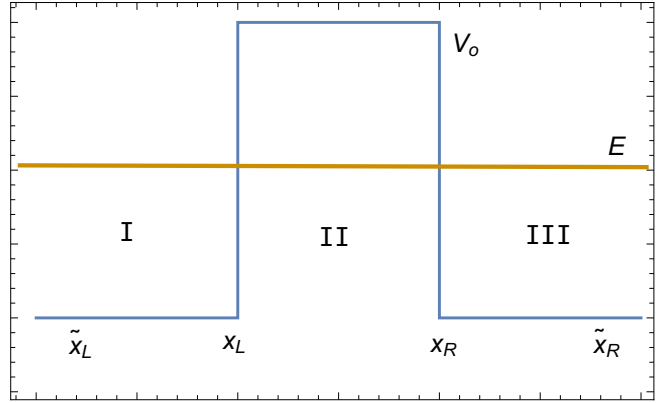}
\caption{The energy diagram for scattering of a particle of energy $E$ from a rectangular potential barrier of height $V_0 > E$ and width $x_R-x_L$. ($\tilde{x}_L$ and $\tilde{x}_R$ are arbitrary points in the regions I and III.)}
\label{figure-1}
\end{figure}

For an insightful analysis of the QTT (\ref{time-2}), it proves convenient to study scattering from a rectangular potential barrier. The setup, illustrated in Fig. \ref{figure-1}, involves  a particle of mass $m$ and energy $E$ incident on a rectangular potential barrier of height $V_0>E$ from the region I ($x< x_L$). It can be reflected back to its incidence region I or it tunnel through region II ($x_L\leq x \leq x_R$) to get to the region III ($x_R < x$). It is the solution of the  Schr{\"o}dinger equation
\begin{eqnarray}
\label{wave-func}
\psi(x)=\left\{\begin{array}{ll} A e^{i k x} + B e^{-i k x}& \;\;\;\;\;\;\;\; ({\rm region\;I})\\
 C e^{-\kappa x} + D e^{\kappa x} & \;\;\;\;\;\;\;\; ({\rm region\;II})\\
e^{i k x}& \;\;\;\;\;\;\;\; ({\rm region\;III})\end{array}\right.
\end{eqnarray}
that determines what alternative is realized with what probability. The probability and probability current remain continuous across $x=x_L$ and $x=x_R$ provided that 
\begin{eqnarray}
\label{coefs}
A &=& \left\{\frac{(k^2-\kappa^2)}{2 i k \kappa} \sinh\left(\kappa (x_R-x_L)\right) +  \cosh \left(\kappa (x_R-x_L)\right)\right\} e^{i k (x_R-x_L)} \nonumber\\
B &=& \frac{(k^2+\kappa^2)}{2 i k \kappa} \sinh\left(\kappa (x_R-x_L)\right)e^{i k (x_R + x_L)}\nonumber\\
C &=& \frac{(-i k + \kappa)}{2 \kappa} e^{(ik + \kappa)x_R}\nonumber\\
D &=& \frac{(i k + \kappa)}{2 \kappa}e^{(i k  - \kappa)x_R}
\end{eqnarray}
where $\hbar k = \sqrt{2 m E}$ ($\hbar\kappa = \sqrt{2m (V_0-E)}$) are inside (outside) momenta, with $V_0>E$.

The wavefunction (\ref{wave-func}),  whose integration constants are fixed in (\ref{coefs}), contains all the information needed for determining physical quantities. For instance, the reflection coefficient
\begin{eqnarray}
R=\frac{|B|^2}{|A|^2}=\frac{\frac{(k^2+\kappa^2)^2}{4 k^2 \kappa^2} \sinh^2(\kappa(x_R-x_L))}{1+\frac{(k^2+\kappa^2)^2}{4 k^2 \kappa^2} \sinh^2(\kappa(x_R-x_L))}
\end{eqnarray}
tends to $1$ ($0$) as $V_0\rightarrow \infty$ $(0)$. The transmission coefficient $T=1-R$ behaves complementarity. The probability current along $\rightsquigarrow\, \equiv\! +\hat{x}$ direction flows as
\begin{eqnarray}
\label{current}
J_{+\hat{x}} = \left\{\begin{array}{ll} \frac{\hbar k}{m}\left(|A|^2 + |A||B|\cos(2 k x + \varphi_{AB})\right)  & ({\rm region\; I})\\
- \frac{\hbar \kappa}{m} |C| |D| \sin\varphi_{CD} & ({\rm region\; II})\\
 \frac{\hbar k}{m}& ({\rm region\; III}) \end{array}\right.
\end{eqnarray}
with the corresponding probability density  
\begin{eqnarray}
\label{density}
\rho-\rho_{-\hat{x}} = \left\{\begin{array}{ll} |A|^2 +  2 |A| |B| \cos(2 k x + \varphi_{AB})  & ({\rm region\; I})\\
  |C|^2 e^{-2\kappa x}  + 2 |C| |D| \cos\varphi_{CD} &({\rm region\; II})\\
1& ({\rm region\; III}) \end{array}\right.
\end{eqnarray}
after defining 
\begin{eqnarray}
\varphi_{AB} &\equiv&  = {Arg}[A B^{\star}]= -2 k x_L - \arctan\left[\frac{2 k \kappa}{k^2 - \kappa^2} \coth{\left(k (x_R - x_L)\right)}\right]\nonumber\\
\varphi_{CD}&\equiv& = {Arg}[C D^{\star}] = -\arctan{\left(\frac{2 k \kappa}{k^2 - \kappa^2} \right)}
\end{eqnarray}
as follows from (\ref{coefs}). 

\begin{figure}[ht]
\centering
\includegraphics[width=8.6cm]{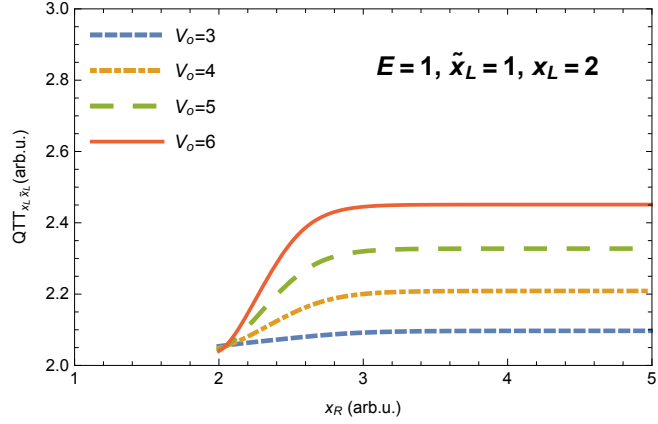}
\caption{
The travel time, namely $({\rm QTT})_{x_L{{\tilde{x}}_L}}$, as a function of the barrier height $V_0$ and barrier width $x_R-x_L$ with ${{\tilde{x}}_L}=1$ and $x_L=2$ units  for the rectangular potential barrier in Fig. \ref{figure-1}. This plot reveals the role of reflection on travel time in region I (larger the $R$ closer the travel time to the limit in (\ref{I:R=1})).}
\label{fig:rectangulartunnel1}
\end{figure}

Having determined probability and probability current, our time formula (\ref{time-2}), the QTT, enables one to determine time it takes to get from  $x=a$ to $x=b>a$ in any region in Fig. \ref{figure-1}. It holds everywhere, inside and outside the barrier. In region I, for instance, time to get form ${\tilde{x}}_L$ to $x_L$ is found to be 
\begin{eqnarray}
({\rm QTT})_{x_L{{\tilde{x}}_L}} = \int_{{\tilde{x}}_L}^{x_L} \frac{\rho^{(I)} -  \rho^{(I)}_{-\hat{x}}}{J^{(I)}_{+\hat{x}}} dx = \frac{2 m (x_L-{\tilde{x}_L})}{\sqrt{2 m E}} - \frac{\hbar}{2 E} \frac{1}{\sqrt{1-R}} \left(\arctan Q(x_L) - \arctan Q({\tilde{x}}_L) \right)
\end{eqnarray}
where  $Q(x)$ is defined as $\sqrt{1+\sqrt{R}}Q(x) = \sqrt{1-\sqrt{R}} \tan \left( k x + \frac{1}{2}\varphi_{AB}\right)$. The time elapsed is seen to depend explicitly on the  reflected wave. For $R=0$ it reduces to
\begin{eqnarray}
\label{I:R=0}
({\rm QTT})_{x_L{{\tilde{x}}_L}} = \frac{m (x_L - {\tilde{x}_L})}{\sqrt{2 m E}}
\end{eqnarray}
which agrees with what is expected of a corpuscular motion and determines. For total reflection ($R=1$), however, it takes a different form 
\begin{eqnarray}
\label{I:R=1}
({\rm QTT})_{x_L{{\tilde{x}}_L}} = 
\frac{2 m (x_L-{\tilde{x}_L})}{\sqrt{2 m E}} - \frac{\hbar}{4 E} \left(\tan\theta(x_L)-\tan\theta({\tilde{x}}_L)\right)
\end{eqnarray}
with $\theta(x)=k(x + \varphi_{AB}/2)$ and significantly differs from time expected of a corpuscular motion. In fact, the QTT in region I is illustrated in Fig. \ref{fig:rectangulartunnel1} as a function of the barrier height and barrier width. It is clear that reflected wave, through the interference term in $\rho$, reverses $\rightsquigarrow$ motion to $\leftsquigarrow$ motion, making the particle to travel around. The figure agrees with the $R=1$ limit in (\ref{I:R=1}). 

\begin{figure}[ht]
\centering
\includegraphics[width=8.6cm]{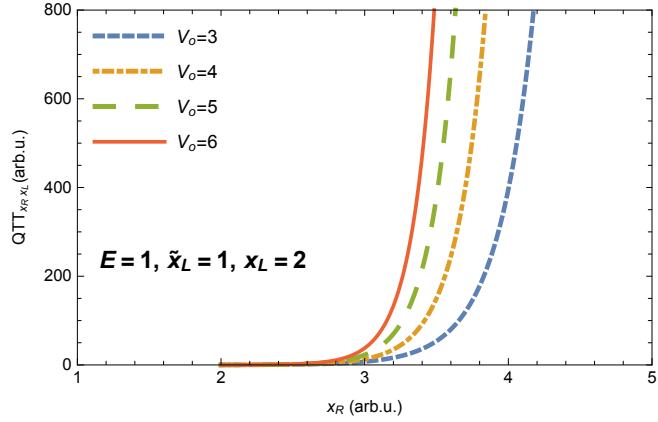}
\caption{The tunneling time, namely $({\rm QTT})_{x_Rx_L}$, as a function of the barrier height $V_0$ and barrier width $x_R-x_L$ with $x_L=2$ units for the rectangular potential barrier in Fig. \ref{figure-1}.}
\label{fig:rectangulartunnel}
\end{figure}

Time it takes to get from $x_L$ and $x_R$ in region II is determined by using the probability current (\ref{current}) and probability density (\ref{density}) in the time formula (\ref{time-2}) so that
\begin{eqnarray}
({\rm QTT})_{{x_R}{x_L}} = \int_{x_L}^{x_R} \frac{\left(\rho^{(II)} -  \rho^{(II)}_{-\hat{x}} \right)}{J^{(II)}_{+\hat{x}}} dx = \frac{m (x_R-x_L)}{\sqrt{2 m E}}\frac{V_0-2E}{V_0-E}
+\frac{\hbar V_0}{8 \sqrt{E(V_0-E)^3}}
\sinh\left(\kappa(x_R-x_L)\right) e^{\kappa(x_R-x_L)}
\label{GTT-II-rect}
\end{eqnarray}
whose physical consistency is justified by the fact that $({\rm QTT})_{{x_R}{x_L}}\rightarrow \infty$ as $V_0\rightarrow \infty$ as well as $x_R-x_L \rightarrow \infty$. These limits are confirmed by  Fig. \ref{fig:rectangulartunnel}, which depicts $({\rm QTT})_{{x_R}{x_L}}$ as a function of the barrier height  and barrier width (in units of $x_R$ with $x_L=0$). It is thus manifest that QTT is physical as increases exponentially with the increasing barrier width and height.
 
Time it takes to get from $x_R$ to ${\tilde{x}}_R$ in region III is given by 
\begin{eqnarray}
({\rm QTT})_{{{\tilde{x}}_R}x_R} = \frac{m ( {\tilde{x}_R-x_R})}{\sqrt{2 m E}}
\end{eqnarray}
as follows from (\ref{I:R=0}) after replacing  $x_L-{\tilde{x}_L}$ with ${\tilde{x}_R}- x_R$. It is due to the absence of any reflected wave in region III that $({\rm QTT})_{{{\tilde{x}}_R}x_R}$ turns out to be precisely what is expected of a corpuscular motion.

Before closing this section, it proves convenient to compare QTT with known time definitions in the literature (for the specific case of the rectangular barrier in Fig. \ref{figure-1}). There exist various time definitions in the literature. The three of them, the entropic tunneling time of \cite{demir-guner}, the phase time of \cite{wigner} and the dwell time of \cite{larmour} are based solely on the potential $V(x)$, and can therefore be unambiguously contrasted for the rectangular potential barrier in Fig. \ref{figure-1}. The entropic tunneling time is based on a statistical approach and holds only in the tunneling region \cite{demir-guner}.  The phase time \cite{wigner,wigner2,wigner3} 
\begin{eqnarray}
\label{phase-time}
(\Delta t)_{\rm phase} = \delta t {\rm (phase\ shift\ in\ wave packet\ peak)} + \frac{m(x_R-x_L)}{\sqrt{2 m E}}\ {\rm (``added\ by\ hand")}
\end{eqnarray}
is composed of the delay from the phase shift in the peak of the wave packet (and have been much disputed due to the absence of a wave packet peak in the tunneling region \cite{time-review}) and the time it would take to traverse $x_R-x_L$ in the absence of barrier via the classical motion (this piece is added by hand not formulated). Obviously, the  QTT produces this "added-by-hand" piece naturally in the regions I and III, and shows that there is no such thing in region II.  The dwell time,  already discussed below equation (\ref{time-2}),  concerns how long the particle stays in a domain. The entropic, phase and dwell times possess the asymptotic limits
\begin{eqnarray}
\label{unphys}
\left(\Delta t\right)_{\rm entropic} &\rightarrow& \infty\\ \left(\Delta t\right)_{\rm phase} &\rightarrow& \infty\\
\left(\Delta t\right)_{\rm dwell}&\rightarrow& \frac{\hbar }{V_0} \sqrt{\frac{E}{V_0-E}}
\end{eqnarray}
as $x_R-x_L \rightarrow \infty$. It is clear that the entropic tunneling time, holding only in the tunneling region, remains subluminal, as also confirmed by Fig. \ref{fig:rectangulartunnel}. The phase time, too, remains subluminal but this happens thanks to the added-by-hand classical time in (\ref{phase-time}).  The phase time suffers from superluminality and, more strikingly, the finite value it takes as $x_R-x_L \rightarrow \infty$ vanishes as $V_{0} \rightarrow \infty$, meaning that the particle tunnels through an infinitely wide and high potential barrier instantaneously. This effect, the Hartman effect \cite{superluminal}, renders the dwell time unphysical. 

In summary, the QTT holds in all three regions I, II and III and leads to  physically sensible results.  What remains is to contrast it with experiment, and that we do in the next section.

\begin{table*}
\caption{\label{tab:const}The various parameters entering the effective potential $V_{eff}(\eta)$ in (\ref{eq:veff}). }
\centering
\begin{tabular}{llllllll}
 Atom&Z&A&B
&C&$I_p^{0}(a.u.)$&$\alpha_N$&$\alpha_I$\\ \hline
 He&2&0&0&2.134&0.903&1.38&0.28 \\
 Ar&18&5.4&1&3.682&0.580&11.1&7.20 \\
 Kr&36&6.42&0.905&4.2&0.515&16.7&9.25\\
\end{tabular}
\end{table*}

\section{QTT In Tunnel Ionization of Atoms}
In this section, we contrast QTT with the formation duration of the  tunneling-enabled $\rm He^+$, $\rm Ar^+$ and $\rm Kr^+$ ions \cite{Maurer,Yakaboylu}.

In the setup \cite{Yakaboylu}, laser pulse propagates along positive $z$ axis, with the electric field 
\begin{equation}
\label{laser-field}
\mathbf{E}(t)=-E_0 f(t) [\cos(\omega t)\hat{\mathbf{x}}+\zeta\sin(\omega t)\hat{\mathbf{y}}]
\end{equation}
such that $f(t)=\cos^4(t\pi/\tau)$,  $E_0=\sqrt{I(1+\zeta^2)}$, $\omega=0.035$, $\tau=156\, {\rm fs}$, and  $\zeta=0.85$. Here $\tau$ is the period of the laser field (much larger than tunneling duration) and $\zeta$ sets the elliptic polarization of the electric field. It is clear that for times $t \ll \frac{\pi\tau}{2}$ the envelope $f(t)$ remains practically unity (static electric field) and stationarity of tunneling is ensured. 

The single-electron effective potential (in atomic units) has the form 
\begin{equation}
\label{pot}
V(\mathbf{r})=-\frac{1}{r}-\frac{\Phi(r)}{r}
\end{equation}
wherein the SAE correction term
\begin{equation}
\Phi(r)=A\exp(-Br)+(Z-1-A)\exp(-Cr)
\end{equation}
has been computed in \cite{Muller,Cloux,Zhang}, with atomic number $Z$ and empirical parameters $A,B,C$ tabulated in Table~\ref{tab:const} for each of $\rm He$, $\rm Ar$ and $\rm Kr$.

In addition to $\Phi$ in (\ref{pot}), there arise further corrections to due to the polarization of the atom under the electric field (basically it becomes an electric dipole). At large distances\cite{Dimitrovski}, the potential then takes the form
\begin{equation}
V(\mathbf{r,E})=-\frac{1}{r}-\frac{\Phi(r)}{r}-\alpha_I\frac{\mathbf{r}\cdot\mathbf{E}}{r^3}+\mathbf{r}\cdot\mathbf{E}\label{eq:pot}
\end{equation}
where $\alpha_I$ is the polarizability of the $\rm He^+$, $\rm Ar^+$ and $\rm Kr^+$ ions.

The ionization energy of polarized atom shifts in proportion with the square of electric field (Stark effect)
\begin{equation}
\label{ion-ener}
I_p=I_p^0+\frac{1}{2}(\alpha_N-\alpha_I)E_0^2
\end{equation}
where $\alpha_N$ is the static polarizability, and $I_p$ and $I_p^0$ are perturbed and unperturbed energies, respectively. They are included in Table~\ref{tab:const}.

Using the potential (\ref{eq:pot}) and the ionization energy  (\ref{ion-ener}) the Schr{\"o}dinger equation takes the form 
\begin{equation}
-I_p\psi(\mathbf{r})=\left(-\frac{\nabla^2}{2}-\frac{1}{r}-\frac{\Phi(r)}{r^3}+\alpha_I\frac{xE_0}{r}-xE_0\right)\psi(\mathbf{r})\label{eq:tise}
\end{equation}
which is known to be separable in the parabolic coordinates \cite{Landau} defined as $\eta=r+x$,  $\xi=r-x$ and  $\phi=\arctan(z/y)$ so that $r=\sqrt{x^2+y^2+z^2}=(\eta+\xi)/2$. 

\begin{figure}[hbt!]
\centering
\includegraphics[width=8.6cm]{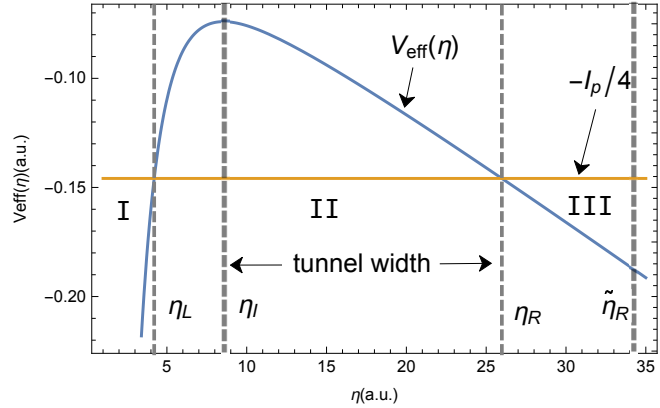}
\caption{The energy diagram of the one-dimensional Schr{\"o}dinger equation (\ref{eq:newtise}).}
\label{figure-2}
\end{figure}

The Schr{\"o}dinger equation (\ref{eq:tise}) transmutes to  one dimension (along $\eta$)
\begin{equation}
-\frac{I_p}{4}N(\eta)=-\frac{1}{2}\frac{\partial^2N(\eta)}{\partial\eta^2}+V_{eff}(\eta)N(\eta)\label{eq:newtise},
\end{equation}
with the effective potential
\begin{eqnarray}
V_{eff}(\eta)=-\frac{1}{8\eta^2}-\frac{1}{2\eta}-\frac{\Phi(\eta/2)}{2\eta} + \alpha_I\frac{E_0}{\eta^2}-\frac{E_0\eta}{8}+\frac{\sqrt{2I_p}}{4\eta}\label{eq:veff}
\end{eqnarray}
arising in the limit $\eta\gg\xi$ ($\eta\simeq 2x \simeq 2r$) \cite{Bisgaard} after separating the wavefunction in the form  $\psi(\eta,\xi,\phi)=\frac{N(\eta)}{\sqrt{\eta}}\frac{X(\xi)}{\sqrt{\xi}}F(\phi)$. The effective potential is depicted schematically in Fig. \ref{figure-2}. 

Hereon, the one-dimensional Schr{\"o}dinger equation (\ref{eq:newtise}) is the topic of investigation.  It should give a satisfactory description of atomic ionization (single electron tunneling) when the electron leaving the atom assumes negligible transverse motion (in $\xi$ and $\phi$ directions) \cite{Maurer,Yakaboylu}.  

For smooth potentials like (\ref{eq:veff}) (see Fig. \ref{figure-2} where $V_{eff}(\eta)$ is depicted) the Schr{\"o}dinger equation (\ref{eq:newtise}) admits a piece-wise WKB solution 
\begin{eqnarray}
\label{eq:wave-func}
N(\eta)=\left\{\begin{array}{ll} \frac{A}{\sqrt{k(\eta)}} e^{i\int_{\eta_L}^\eta  \! k(\eta') \, \mathrm{d}\eta'} + \frac{B}{\sqrt{k(\eta)}} e^{-i\int_{\eta_L}^\eta  \! k(\eta') \, \mathrm{d}\eta'}&\;\;(\rm region\; I)\\
 \frac{C}{\sqrt{\kappa(\eta)}} e^{\int_{\eta_R}^\eta  \! \kappa(\eta') \, \mathrm{d}\eta'} + \frac{D}{\sqrt{\kappa(\eta)}} e^{-\int_{\eta_R}^\eta  \! \kappa(\eta') \, \mathrm{d}\eta'} &\;\;(\rm region\; II)\\
\frac{1}{\sqrt{k(\eta)}}e^{i\int_{\eta_R}^\eta  \! k(\eta') \, \mathrm{d}\eta'}&\;\;(\rm region\; III)\end{array}\right.
\end{eqnarray}
in which
\begin{eqnarray}
k(\eta)=\left(2\left(-\frac{I_p}{4}-V_{eff}(\eta)\right)\right)^{1/2}
\end{eqnarray}
and
\begin{eqnarray}
\kappa(\eta)=\left(2\left(V_{eff}(\eta)+\frac{I_p}{4}\right)\right)^{1/2}
\end{eqnarray}
are the momenta outside and inside the tunneling region, respectively. After determining the constants $A,\dots, D$ by patching the wavefunctions at the turning points, the wavefunction $N(\eta)$ take the definitive form (in the same regions as in (\ref{eq:wave-func}))
\begin{eqnarray}
\label{eq:wave-final}
N(\eta)=\left\{\begin{array}{l} \frac{i}{\sqrt{k(\eta)}} \left\{\left(e^{\chi}+\frac{1}{4} e^{-\chi} \right) e^{i\int_{\eta_L}^\eta  \! k(\eta') \, \mathrm{d}\eta'} + i \left(-e^{\chi}+\frac{1}{4} e^{-\chi} \right) e^{-i\int_{\eta_L}^\eta  \! k(\eta') \, d\eta'} \right\}\\
 \frac{i}{\sqrt{\kappa(\eta)}} \left\{\frac{\sqrt{i}}{2}e^{\int_{\eta_R}^\eta  \! \kappa(\eta') \, \mathrm{d}\eta'} + \frac{1}{\sqrt{i}} e^{-\int_{\eta_R}^\eta  \! \kappa(\eta') \, d\eta'} \right\}\\
\frac{i}{\sqrt{k(\eta)}} e^{i\int_{\eta_R}^\eta  \! k(\eta') \, d\eta'}
\end{array}\right.
\end{eqnarray}
where $\chi=\int_{\eta_L}^{\eta_R}  \kappa(\eta) d\eta$ is the total phase accumulated as the electron tunnels from one side of the barrier to the other. 

\begin{figure}[hbt!]
\centering
\includegraphics[width=8.6cm]{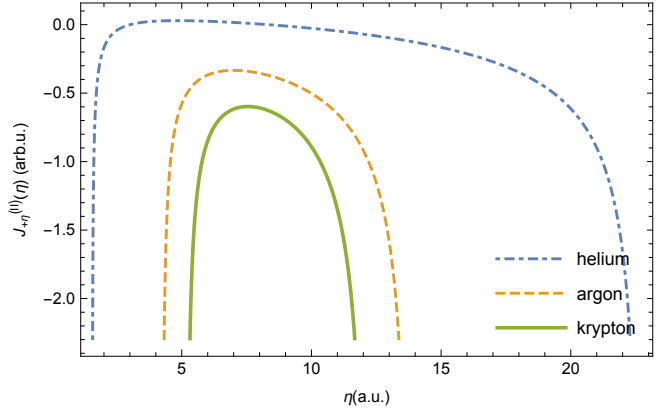}
\caption{\label{fig:currentr} The probability current density ${J_{+\hat{\eta}}}$ (in region II of the $V_{eff}(\eta)$ in Fig. \ref{figure-2}) for a laser intensity of $I=3.6\times10^{14} W/cm^2$.}
\end{figure}

We now use the wavefunction (\ref{eq:wave-final}) in the time formula (\ref{time-2}) to determine the QTT in any region (I, II or III) of the potential landscape. The first thing one must notice about the WKB wavefunction is that 
\begin{eqnarray}
\left|N^{\rm (II)}(\eta)\right|^2 = \frac{1}{\kappa(\eta)} \left(\frac{1}{4} e^{2\int_{\eta_R}^\eta \kappa(\eta') d\eta'} + e^{-2\int_{\eta_R}^\eta  \kappa(\eta') d\eta'}\right)
\end{eqnarray}
namely there exists no interference (more correctly, {\it overlap})  between the forward and backward propagating waves inside the tunneling region. This means that the particle entering from $x_L$ proceeds en route to $x_R$, with no possibility of derailing to a backward propagating wave (as was also utilized in  \cite{demir-guner}).  The two probabilities are disjoint with no overlap \cite{Sokolovski,buttiker-landauer3}. This enables us to define right and left currents explicitly and calculate travel time QTT straightforwardly. In fact, using the time formula (\ref{time-2}) we define tunneling time through the potential barrier in Fig. \ref{figure-2} as  
\begin{eqnarray}
({\rm QTT})_{{\eta_R}{\eta_I}} &=& \int_{\eta_I}^{\eta_R} \frac{\left(\rho^{(II)} -  \rho^{(II)}_{-\hat{\eta}} \right)}{J^{(II)}_{+\hat{\eta}}} d\eta
\label{eq:time}
\end{eqnarray}
where the initial point  $\eta_I$ and the final point $\eta_R$ (the right turning point) are listed in Table~\ref{tab:positions} for two values of the laser intensity.In the region II of the potential barrier in Fig. \ref{figure-2} one finds
\begin{eqnarray}
\rho^{(II)} -  \rho^{(II)}_{-\hat{\eta}} = \frac{1}{4\kappa(\eta)}e^{2\int_{\eta_R}^\eta  \! \kappa(\eta') \, \mathrm{d}\eta'}
\label{eq:density}
\end{eqnarray}
for probability density, and 
\begin{eqnarray}
{J^{(II)}_{+\hat{\eta}}} = \frac{1}{2}-\frac{\kappa^{'}(\eta)}{4\kappa^2(\eta)}
\label{eq:current}
\end{eqnarray}
for probability current along $+\hat{\eta}$ direction. 


The energy configuration behind the Schr{\"o}dinger equation (\ref{eq:newtise}) is depicted in Fig. \ref{figure-2}.The three regions I, II, III, having a similar meaning as those in Fig. \ref{figure-1}, correspond to three different propagation regimes for the electron. The region I, overwhelmed by the Coulombic singularity at the position of the nucleus, is skipped by the tunneling dynamics in that $V_{eff}(\eta)$ attains its maximum when the laser intensity is maximum (corresponding to $t=0$ at which electric field strength equals $E_0$). For this reason, time calculation starts not with $\eta_L$ but with $\eta_I$ in Fig. \ref{figure-2} at which $V_{eff}(\eta)$ is maximum (or current density is maximum). To see this, we plot in Fig. \ref{fig:currentr} the probability current density 
${J_{+\hat{\eta}}}$ in (\ref{eq:current}). It is clear that maximum of the potential ($\eta=\eta_I$) changes from atom to atom, with the obvious fact that the region I remains tiny for each atom.

\begin{figure}[hbt!]
\centering
\includegraphics[width=8.6cm]{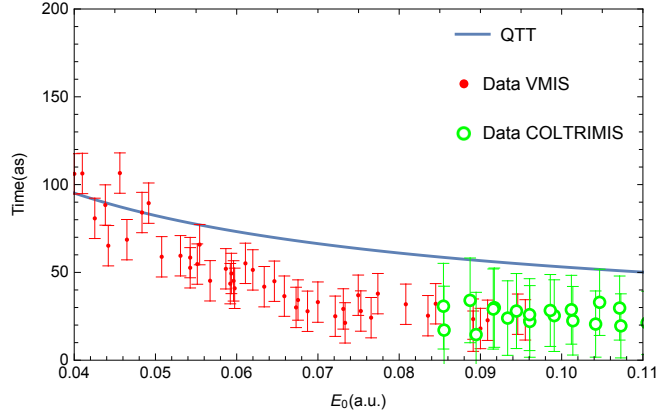}
\caption{\label{fig:hetime-deney} The $\rm He$ ionization time as modelled by $({\rm QTT})_{{\eta_R}{\eta_I}}$ and as measured by the VMIS and COLTRIMS spectrometers \cite{Maurer}.}
\end{figure}

The tunneling time measurements are based on ionization of atomic electrons through laser-controlled potential barriers. The intense laser pulse (around $10^{14}\ {\rm W/cm^2}$) turns the Coulomb potential into a potential barrier as in Fig. \ref{figure-2} so that a valence electron can tunnel to continuum, in which it continues to be accelerated by the laser field which itself varies with time as in (\ref{laser-field}). The measurement of the $\rm He$ ionization time in \cite{Maurer} is based on momentum distribution of the continuum electrons (as measured by COLTRIMS and VMIS spectrometers). The technique involves extraction of the phase of the electric field at the tunnel exit ($\eta_R$ in Fig. \ref{figure-2}) from electron's momentum distribution, and takes into account Coulombic and drift corrections. The main stage of the extraction is the determination of the phase angle that leads to the wavepacket whose peak gives the most probable electron trajectory \cite{Maurer} (see also the recent review \cite{Maurer2}). This $\rm He$ ionization data can be used to determine how realistic the QTT is. The tunnel exit should be taken near the fall of probability density current, for a proper comparison. The results are shown in Fig. \ref{fig:hetime-deney} where superimpose $({\rm QTT})_{{\eta_R}{\eta_I}}$ in (\ref{eq:time}) on the experimental data (Fig 3 (b) of \cite{Maurer}). It is clear that there is good agreement between the QTT and the experimental data. The slight difference between the two is expected on the grounds that modeling of the $\rm He$ ionization are different in the experiment of \cite{Maurer} and in the Schr{\"o}dinger equation (\ref{eq:newtise}). The experiment, to our understanding, sets the ionization energy as $I^{(exp)}_p=(\gamma E_0 \omega^{-1})^2/2$ (the Keldysh parameter $\gamma\sim 0.8 - 2.5$ controls the tunnel ionization regime \cite{keldysh,keldysh2}) and determines tunneling time transition of electron from $\eta_L$ to $\eta_R$ with that $I_p=I^{(exp)}_p$. Atomically, however, the ionization energy $I_p^0$ is precisely known and improved ionization energy $I_p$ involves $E_0$ via only the correction terms in (\ref{ion-ener}). Moreover, tunneling dynamics can be known and ensured only beyond $\eta_I$ (not $\eta_L$) \cite{Yakaboylu}, and that is the point used in $({\rm QTT})_{{\eta_R}{\eta_I}}$ in (\ref{eq:time}). Despite there differences, however, QTT and the experimental data show good agreement and, with increasing experimental and theoretical precision (going beyond the WKB, for instance) the closeness in  Fig. \ref{fig:hetime-deney} can turn into a complete agreement.

\begin{figure}[hbt!]
\centering
\includegraphics[width=8.6cm]{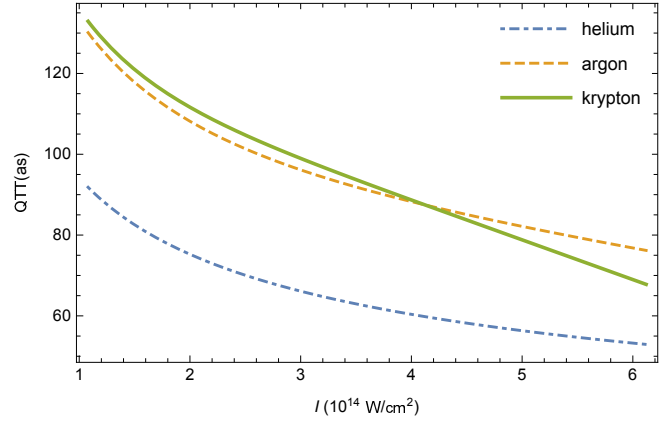}
\caption{\label{fig:timeintensity} QTT as a function of the laser intensity $I$ (electric field strength $E_0$ ranges from 0.04 to 0.1 au).}
\label{fig-GTT}
\end{figure} 

The comparative measurement of the $\rm Kr$ and $\rm Ar$ ionization times in \cite{Yakaboylu}, which is an efficient method for  eliminating various systematic and statistical effects, focuses on the continuum electron trajectories. The continuum electrons are assumed to follow classical laws of motion (in contrast to \cite{Maurer}, which assumes a wavepacket), and electron trajectories are found to explain the data if there is a time delay in the potential barrier. They adopted phase time \cite{wigner,wigner2} in their analysis. Our approach differs form theirs in two aspects:
\begin{enumerate}
    \item We use the QTT in the tunneling region, and
    \item We continue to use the QTT in the continuum.
\end{enumerate}
In the tunneling region, for $\rm Kr$ ionization, the phase time formalism gives 138 {\rm as} at  $I =1.08\times10^{14}\ {\rm W/cm^2}$, 126 {\rm as} at  $I =1.7\times10^{14}\ {\rm W/cm^2}$, and 64 {\rm as} at  $I =6.12\times10^{14}\ {\rm W/cm^2}$ (see Fig. 2(b) of \cite{Yakaboylu}). For the same laser intensities, the QTT is found to be 133, 116 and 68 {\rm as}, respectively (see Fig.\ref{fig:timeintensity}). The two times remain close to each other throughout.  Here, it should be noted that phase time pertains to the dominant path along the tunneling channel but the time within the tunneling region is pure imaginary (see the discussions in \cite{time-review,demir-guner}). This feature ensures that the Wigner time in \cite{Yakaboylu} is dominantly the one coming from the overall real-time phase of the propagator, and corresponds to the time added by hand in (\ref{phase-time}). This feature is what makes time to grow almost linearly with the barrier width (see Fig. 2(b) of \cite{Yakaboylu}). In essence, therefore, the closeness between the QTT and the phase time results from mainly the overall phase of the kernel corresponding to the added-by-hand time amount in (\ref{phase-time}).

The $({\rm QTT})_{{\eta_R}{\eta_I}}$ curves in Fig. (\ref{fig:timeintensity}), plotted for each of  $\rm He$, $\rm Kr$ and $\rm Ar$, combines $\rm He$ with the other atoms and give a clear view of how tunneling time changes from atom to atom. The separation between the $\rm He$ and the other two reveals the impact of the effective potential (\ref{eq:veff}) (to the exent it can be applied to low-$Z$ atoms like $\rm He$). The closeness of the $\rm Kr$ and $\rm Ar$ curves, on the other hand, justifies the experimental technique employed in \cite{Yakaboylu} in that various systematic effects can indeed be discarded by experimenting two atoms simultaneously. The drop in the $\rm Kr$ tunneling time at large intensities (related to its dipole structure) agrees with the results of \cite{Yakaboylu}. It may be concluded that  QTT has the potential to be reliably applied to tunnel ionization of different atoms.  

\begin{figure*}[hbt!]
 \centering
  \includegraphics[width=.5\linewidth]{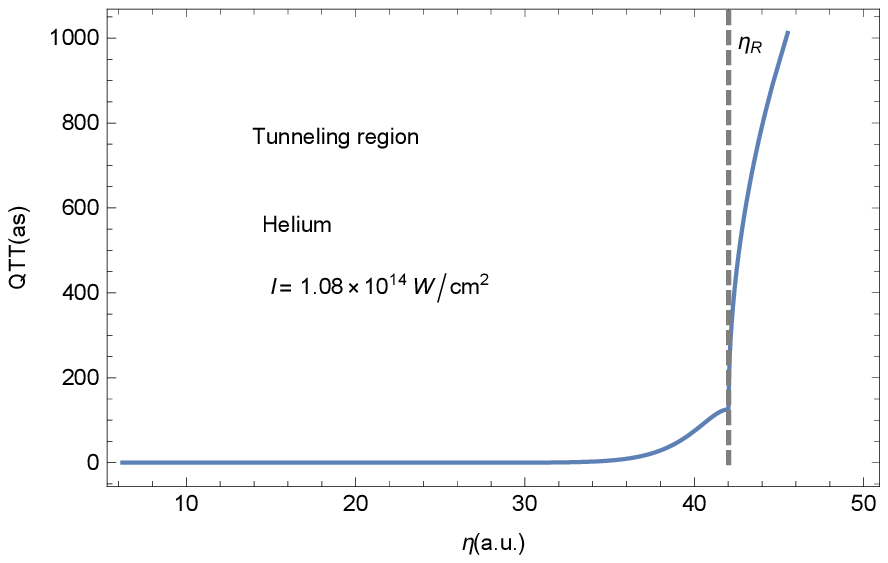}\hfill
  \includegraphics[width=.5\linewidth]{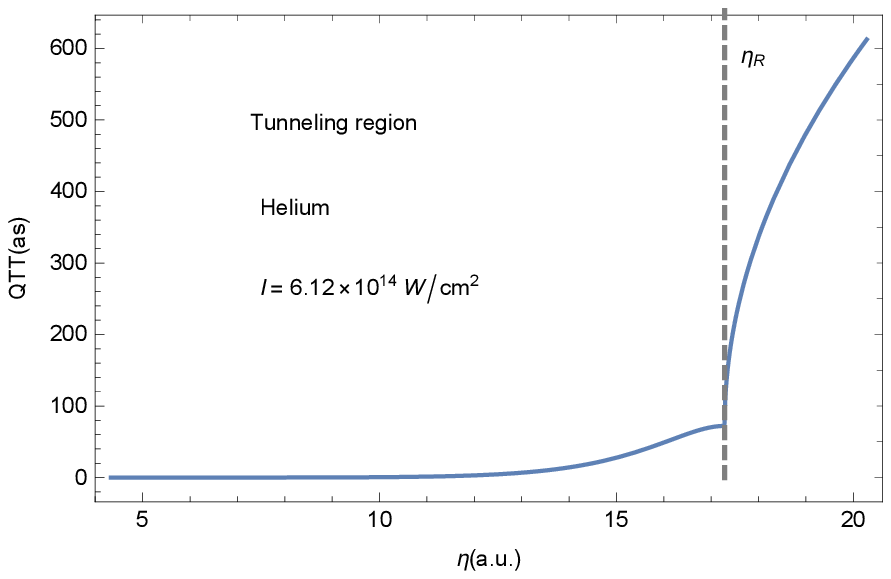}
  \includegraphics[width=.5\linewidth]{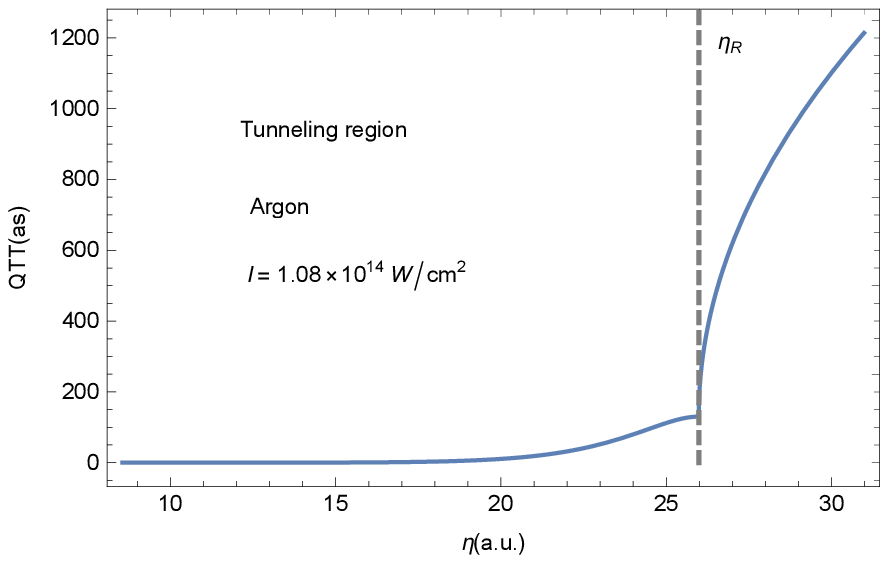}\hfill
  \includegraphics[width=.5\linewidth]{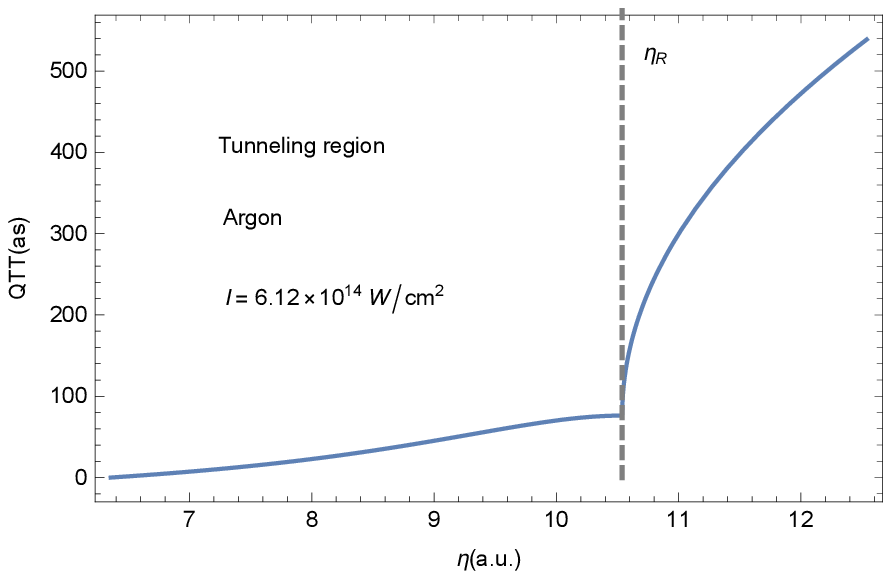}
  \includegraphics[width=.5\linewidth]{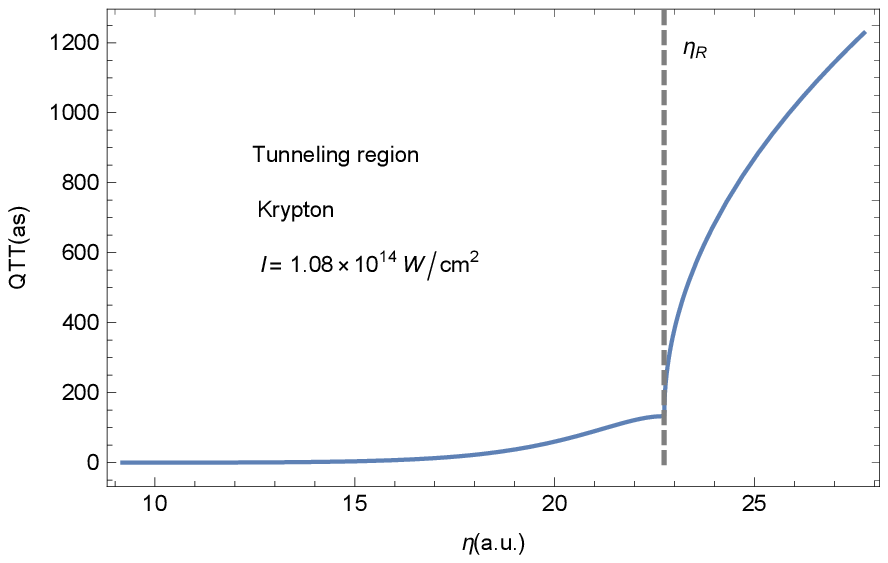}\hfill
  \includegraphics[width=.5\linewidth]{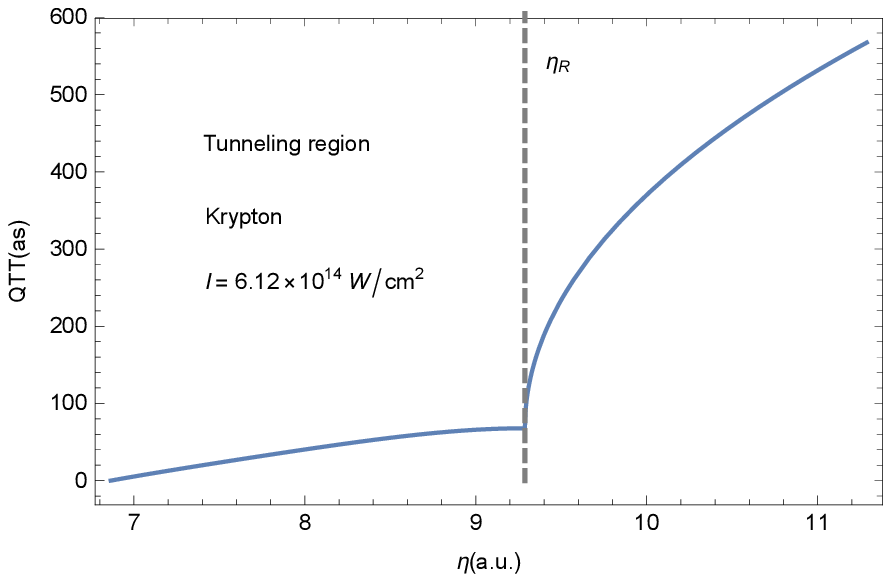}
  \caption{The QTT in the regions II and III of the effective potential in Fig. \ref{figure-2} for $\rm He$ (top), $\rm Ar$ (middle) and $\rm Kr$ (bottom) for  $I=1.08 \times 10^{14}\ {\rm W/cm^2}$ (deep-tunneling on left side) and  $I=6.12 \times 10^{14}\ {\rm W/cm^2}$ (near-threshold-tunneling on right side), where $\eta_R$ is starting point for the continuum electron propagation. The QTT is continuous at $\eta=\eta_R$ but has different slopes in the two sides.}
  \label{fig:time-trajectory}
\end{figure*}
\begin{table}
\caption{\label{tab:positions} The turning points $\eta_L$, $\eta_R$ and maximum point $\eta_I$ for two different laser intensities.}
\centering
\begin{tabular}{cccc}
 Atom&$\eta_L$&$\eta_I$&$\eta_R$\\ \hline
 &&Intensity 1.08$\times 10^{14}\ {\rm W/cm^2}$&\\ \hline
 He&1.5358&6.2307&42.0210 \\
 Ar&4.2036&8.5492&25.9824 \\
 Kr&5.0274&9.1864&22.7422 \\\hline
 &&Intensity 6.12$\times 10^{14}\ {\rm W/cm^2}$&\\ \hline
 He&1.5477&4.3271&17.2830 \\
 Ar&4.2493&6.3563&10.5383 \\
 Kr&5.2817&6.8643&9.2879 \\
\end{tabular}
\end{table}

After traversing the region II (completion of tunneling process), the continuum electron continues to propagate in region III of the potential landscape in Fig. \ref{figure-2}. The scattering process continues to maintain its stationary character for propagation duration sufficiently short compared to the laser period $\tau=156\, {\rm fs}$, and QTT can safely be employed as
\begin{eqnarray}
({\rm QTT})_{{{\tilde \eta}_R}{\eta_R}} &=& \int^{{{\tilde \eta}_R}}_{\eta_R} \frac{\left(\rho^{(III)} -  \rho^{(III)}_{-\hat{\eta}} \right)}{J^{(III)}_{+\hat{\eta}}} d\eta
\label{eq:timeIII}
\end{eqnarray}
 such that
\begin{eqnarray}
\rho^{(III)} -  \rho^{(III)}_{-\hat{\eta}} =\frac{1}{k(\eta)}
\label{eq:density-r}
\end{eqnarray}
and 
\begin{eqnarray}
{J^{(III)}_{+\hat{\eta}}} = 1
\label{eq:current-r}
\end{eqnarray}
where ${\tilde\eta}_R$ is a point in region III. 

Plotted in Fig. \ref{fig:time-trajectory} is the variation of the QTT with $\eta$ in tunneling (left of $\eta_R$ divide) and continuum (right of $\eta_R$ divide) regions for $\rm He$ (top), $\rm Ar$ (middle) and $\rm Kr$ (bottom) atoms for intensities  $I=1.08 \times 10^{14}\ {\rm W/cm^2}$ (deep-tunneling on left side) and  $I=6.12 \times 10^{14}\ {\rm W/cm^2}$ (near-threshold-tunneling on right side). In all panels, $\eta_R$ is the starting point for the continuum electron, and the total time elapsed for reaching the point ${\tilde \eta}_R$ in Fig. \ref{figure-2} depends on the delay within the tunneling region. It is precisely this delay that is measured (comparatively between the $\rm Ar$  and $\rm Kr$ ionizations) in \cite{Yakaboylu}. Our results differ from  calculations in \cite{Yakaboylu} mainly at two points:
\begin{enumerate}
    \item QTT varies non-linearly inside the barrier (it does not follow the dominant $(\eta-\eta_I)/\sqrt{2E}$ linear behavior of the phase time),
    \item QTT grows fast outside the barrier (it asymptotes to $(\eta-\eta_R)/\sqrt{2E}$ behavior at large $\eta$ where $V_{eff}(\eta)$ is diminished).
\end{enumerate}
The first difference above is not hard to make sense, given the behavior of the phase time. The second point is, however, more subtle. Speaking specifically, we have not been able to reproduce Fig. 2 (a) of the \cite{Yakaboylu} even with their parameters; their plots seem to involve some kind of rescaling and shifting. Concerning this difference, assuming the validity of the SAE potential throughout, it should be emphasized that probability conservation ensures continuity of the wavefunction and its first derivative,  and this ensures that QTT flows continuously across a boundary but its derivative (the second derivatives of the wavefunction) does not have to be continuous, and thus, time outside the barrier (region III) can grow with a different rate than the one in the tunneling region (region II).

\section{Conclusion}
In this work we have proposed a new time formula -- the quantum travel time -- which holds everywhere (inside and outside the tunneling region) and which applies directly to stationary processes (systems with conserved energy, like the tunneling process). We analyzed the QTT in rectangular (Sec. III) and atomic (Sec. IV) potential configurations, and shown that the QTT leads to physically acceptable results with good agreement with experimental data (despite the use of WKB approximation).  

We should emphasize that the QTT has the potential to give reliable predictions of time elapsed in quantum systems (including tunneling-driven processes). It can be applied (and tested this way) to various biological (like DNA damage), chemical (like astrochemistry) and physical systems (diodes to the formation of the Universe). Our analysis shows that the QTT can lead to realistic predictions in each of these tunneling-driven phenomena. 

DD thanks A. Landsman for sharing (via C. Hofmann) with us the $\rm He$ data used in Fig. 6. DD thanks also to Department of Physics, {\. I}zmir Institute of Technology where this work was started.

\end{document}